\documentclass[pre,twocolumn,aps,amsmath,showpacs]{revtex4}
\usepackage{graphicx}

\newcommand{\beq}{\begin{equation}}
\newcommand{\eeq}{\end{equation}}
\newcommand{\beqs} {\begin{displaymath}}
\newcommand{\eeqs} {\end{displaymath}}
\newcommand{\beqa} {\begin{eqnarray}}
\newcommand{\eeqa} {\end{eqnarray}}
\newcommand{\beqas} {\begin{eqnarray*}}
\newcommand{\eeqas} {\end{eqnarray*}}

\begin{document}

\title{Segregation induced by inelasticity in
a vibrofluidized granular mixture}
\author{R. Brito$^{1,2}$, H. Enr\'{\i}quez$^{2}$, S. Godoy$^1$, and R.
Soto$^{1,2}$}
\affiliation{[1] Departamento de F\'\i sica, FCFM, Universidad
de Chile. Casilla 487-3, Santiago, Chile.}
\affiliation{[2] Departamento de F\'\i sica Aplicada I and GISC,
Facultad de Ciencias F\'\i sicas, Universidad Complutense. 28040 Madrid, Spain.}

\begin{abstract}
We investigate the segregation of  a dense binary mixture of granular particles that
only differ in their restitution coefficient. The mixture is vertically vibrated
in the presence of gravity.
We find a partial segregation of the species, where most dissipative particles
submerge in the less dissipative ones.  The segregation occurs even if one type
of the particles is elastic.
In order to have a complete description of the system, we study the structure of
the fluid at microscopic scale (few particle diameters). The density and temperature pair
distribution functions show strong enhancements respect the equilibrium ones at the
same density.
In particular, there is an increase in the probability that the more inelastic particles group
together in pairs (microsegregation).
Microscopically the segregation is buoyancy driven, by the appearance of a dense
and cold region around the more inelastic particles.
\end{abstract}


\maketitle

\section{Introduction}

Granular matter, when externally excited, show a series of peculiar
phenomena. One of them is the mixing or segregation that takes place when 
two or more species of different grains are put
together. Depending on the control parameters and the energy injection
mechanisms, grains of different size, shape, mass or mechanical properties can
mix or segregate.

Consider the particular case of a mixture of small grains and one large
intruder. Here the intruder can go up \cite{Rosato} or down \cite{Breu} --the direct and
reverse Brazil nut effects, respectively--, phenomenon that has been studied in many papers (see, e.g.
Ref. \cite{Kudrolli} and references therein).
When both species have similar sizes (but possibly different) we can select few cases
where a variety of
segregation mechanisms and scenario appear \cite{RuizSuarez,Schroter, Khakhar}.
For instance, in Ref. \cite{Mullin}
particles of different masses, radii and restitution coefficients are placed in a dish
which is horizontally vibrated, finding complete segregation. Segregation is also found in the same geometry
when the grains have different friction coefficient with the base \cite{Ciamarra}.
Under horizontal swirling, radial
segregation of particles of different sizes has been observed
\cite{Schnautz}. In avalanches, grains of different shape segregate in
stripes \cite{Makse}; in partially filled rotating drums, axial size
segregation develops \cite{Hill}. In two dimensional systems under gravity, sinusoidally
vibrated, clustering has been observed \cite{King}.
This segregation effect can be modulated by using non-sinusoidal vibration \cite{King2}.

In some of the cases mentioned above the grain species differ on the friction or
restitution coefficient. However few papers have studied segregation when
this is the only difference between grains. One of these cases is
Ref. \cite{Kondic}, where a mixture of spheres
that only differ in friction coefficients (static, dynamic and
rolling) is horizontally vibrated. They find complete mixing --that is, no segregation-- for a flat plate
while segregation is only observed when the plate was slightly inclined.
Therefore, these results contradict the previously mentioned ones.

In a theoretical approach, Ref. \cite{Serero} constructs the hydrodynamic
equations from the Boltzmann equation, finding segregation induced by
inelasticity. The authors explain the phenomenon as a consequence of the
temperature
gradient in the system induced by inelastic collisions, and relate the
concentration gradient with the temperature gradient.
In the same spirit, Ref. \cite{Brey} studies the low density hydrodynamics of a mixture in the so called
tracer limit, i.e. where the concentration of one of the components tends to
zero. Among other results, they find that the temperature ratio of both species
must be a constant. This constant value was already measured by two
experimental groups \cite{Menon,Wildman} in 2D and 3D  respectively and by
computer simulations \cite{Paolotti}. Generalization to high density has been done by Garz\'o \cite{Garzo}.

The goal of this paper is to confirm or deny the existence of segregation
induced by a inelasticity difference and characterize this phenomenon.
The main tool will be Molecular Dynamics computer simulations of two dimensional
systems of a binary mixture kept fluidized by a vibrating base.

The structure of this paper is as follows. In Sect. II we describe the system under
consideration. Section III is devoted to the macroscopic study of the system, in particular
density and temperature profiles.
Section IV present a microscopic study via the pair distribution functions. Section V proposes
a model that possibly explains the segregation. We conclude with Section VI summarizing the results of the paper.

\section{Description of the system}
We study the effect of the difference on restitution coefficients in
 the segregation phenomenon, by  means of Molecular Dynamics simulations of a
bidimensional  granular mixture of two types of particles, named 
A and B. Grains are modeled as smooth Inelastic Hard Disks both having the same
diameter $\sigma$ and mass $m$, but differing on the normal restitution
coefficient that characterizes their inelastic collisions.
The restitution coefficient for A-A collisions is $\alpha_A$, for B-B
collisions is $\alpha_B$. For the interparticle collisions A-B  we have taken
$\alpha_{AB}=(\alpha_A+\alpha_B)/2$. In what follows we will consider
that B are the most inelastic particles ($\alpha_B<\alpha_A$).

We have taken a fixed total number of particles $N_T=N_A+N_B$, changing the
concentration of the B particles.
For the simulations reported in this paper, we have fixed $N_T=680$ disks
and varied $N_B$ from 10 (that can be considered as a tracer limit)
until 160. The disks are placed under the action of a
gravitational acceleration $g$ pointing downward in a rectangular
box of width $L_x=50\sigma$, infinite height, and with the bottom wall
oscillating periodically at high frequency $\omega$ and small amplitude
$A$, with a bi-parabolic waveform \cite{Soto}. Periodic boundary conditions are
used in the horizontal direction, trying to avoid the appearance of convective rolls
by the influence of the walls. Under these conditions, the system reaches a stationary
state with gradients in the vertical direction \cite{Grossman}.

Units are chosen such that $\sigma=1$, $m=1$, and we fix the energy scale
by the wall oscillation, $m(A\omega)^2=1$. Simulations are
performed with $g=0.15$, and $A=0.01\sigma$.

\section{Macroscopic segregation} \label{sec.macroseg}

In order to illustrate the main observed features,  we report results of a simulation
having a small fraction of inelastic particles $N_B=10$, and $\alpha_B=0.7$ and
the rest nearly elastic: $N_A=670$, $\alpha_A=0.98$.

\begin{figure}[htb]
\includegraphics[angle=0,clip=true,width=0.95\columnwidth,
keepaspectratio]{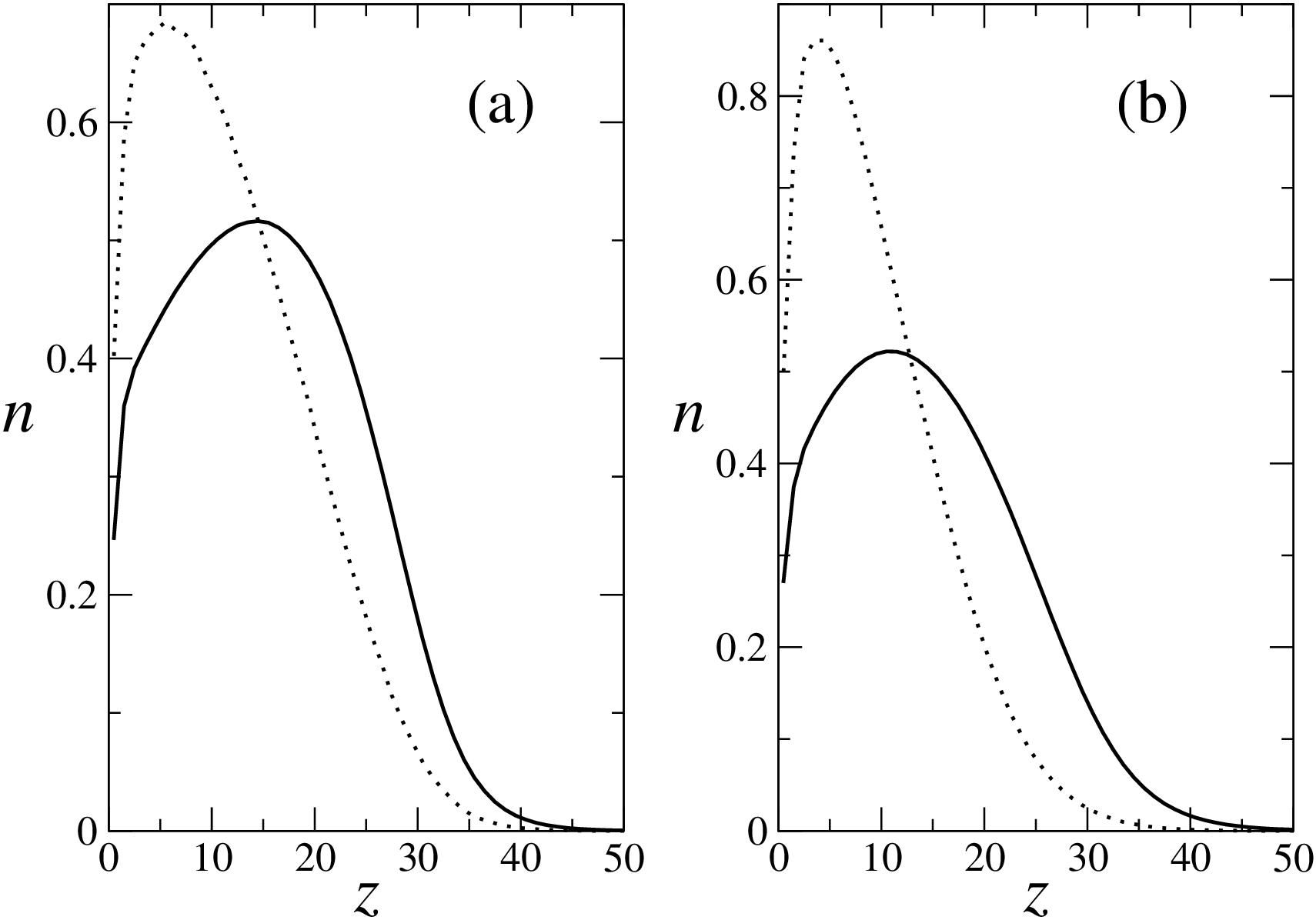}
\caption{Density profiles for two species, density of A, $n_A$, (solid
line) and the rescaled density of B, $n_B N_A/N_B$, (dotted line). (a):
both species are inelastic with $N_A=670$, $N_B=10$, $\alpha_A=0.98$,
$\alpha_B=0.7$. (b): A is elastic while B is inelastic and $N_A=640$,
$N_B=40$, and $\alpha_B=0.5$}
\label{fig.densprofiles}
\end{figure}

The density profiles of the two species are shown in
Fig. \ref{fig.densprofiles}a, where we plot
the number density of particles of type A and B: $n_A(z)$ and $n_B(z)$. The normalization
of these quantities is such that $\int_0^\infty dz \int_0^{Lx}dx \, n_A(z) = N_A$ (resp. B).
For plotting purposes only, $n_B$ is rescaled by a factor $N_A/N_B$, so in the case of no segregation both profiles would be identical.
Both densities have the characteristic shape of
vibrofluidized systems subject to gravity: there is a initial density
increase due to the abrupt temperature drop caused by dissipation, and at
higher positions, density decreases again due to gravity
\cite{Grossman}. Density exhibits a maximum at $z\simeq 15\sigma$
where the density $n\simeq 0.5$, so the system cannot be considered as dilute.
The density profile of the more inelastic particles, B, is plotted as a dotted line in Fig 1a.
Its maximum is located at smaller $z$,  indicating that they are closer to the
bottom of the container as compared to the more elastic ones. Therefore, particles segregate although the segregation is not complete.

\begin{figure}[htb]
\includegraphics[angle=0,clip=true,width=0.95\columnwidth,
keepaspectratio]{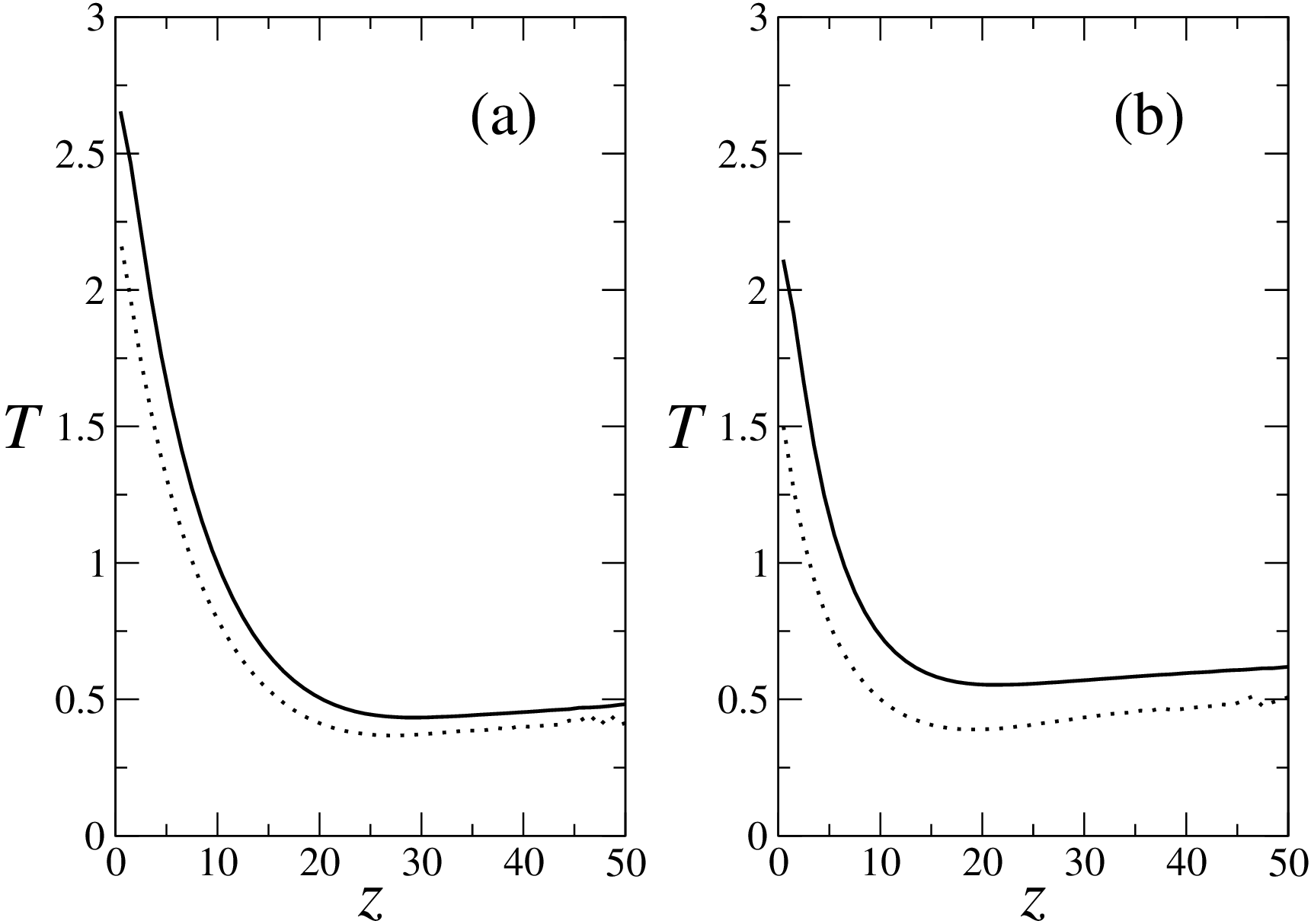}
\caption{Temperature profiles for two species, $T_A$ (solid
line) and $T_B$ (dotted line). (a) and (b) graphics have the same
parameters as in Fig. \ref{fig.densprofiles}}
\label{fig.tempprofiles}
\end{figure}

The temperature profiles are also highly inhomogeneous, as shown in Fig.
\ref{fig.tempprofiles}a.
For both
species, temperature presents a initial abrupt drop, but later (after
$z\simeq 20\sigma$ both profiles present a linear increase with height.
This phenomenon was already observed in one-component systems, and it is
associated to the energy transport term, $-\mu\nabla n$,  associated to density
gradients, that appear in granular fluids \cite{RamirezSoto}. Let us note that the
maximum density does not coincide with the temperature minimum: there is a shift
between these two quantities which is qualitatively described by a hydrostatic
balance in presence of gravity \cite{RamirezSoto,Serero}.

The described segregation of species A and B is produced by their
different restitution coefficients as all the other properties are the same.
To study in more detail the effect of the difference of inelasticities and
to understand the origin of this particular segregation, we proceed to
study the limiting case in which the A particles are elastic
($\alpha_A=1$) and only the B particles are
inelastic (and consequently, collisions A-B are also inelastic).
In this way we also limit the parameter space, allowing to
a more detailed quantitative study.

Figures  \ref{fig.densprofiles}b and
\ref{fig.tempprofiles}b show the density and temperature profiles for such case,
where particles of type A are elastic ($\alpha_A=1$) particles B are inelastic
($\alpha_B=0.5$) considering $N_B=40$.  It is observed
that the main properties of the profiles are preserved, even the positive
slope of $T_A$ despite the A-A collisions are elastic. Partial segregation is again observed,
where inelastic particles, B, sink to the bottom of the container while elastic ones, A,
are majority at upper layers of the fluid.

These results are not surprising in view of the predictions
of Ref.~\cite{Serero}, where it is argued that the segregation is
produced when the particles with different restitution coefficients are immersed
in a temperature gradient. The gradient is
induced by the inelastic collisions, so such gradient can be created vibrating a
mixture of elastic and inelastic particles. The latter ones dissipate some of the energy
injected by vibration creating a stationary state. The hydrodynamic description
of the mixture also contains the dissipative flux $ -\mu \nabla n $, and
therefore it is expected that 
the hydrodynamic profiles of density and temperature will be equivalent to a
full inelastic system.

\begin{figure}[htb]
\includegraphics[clip=true,width=0.9\columnwidth,
keepaspectratio]{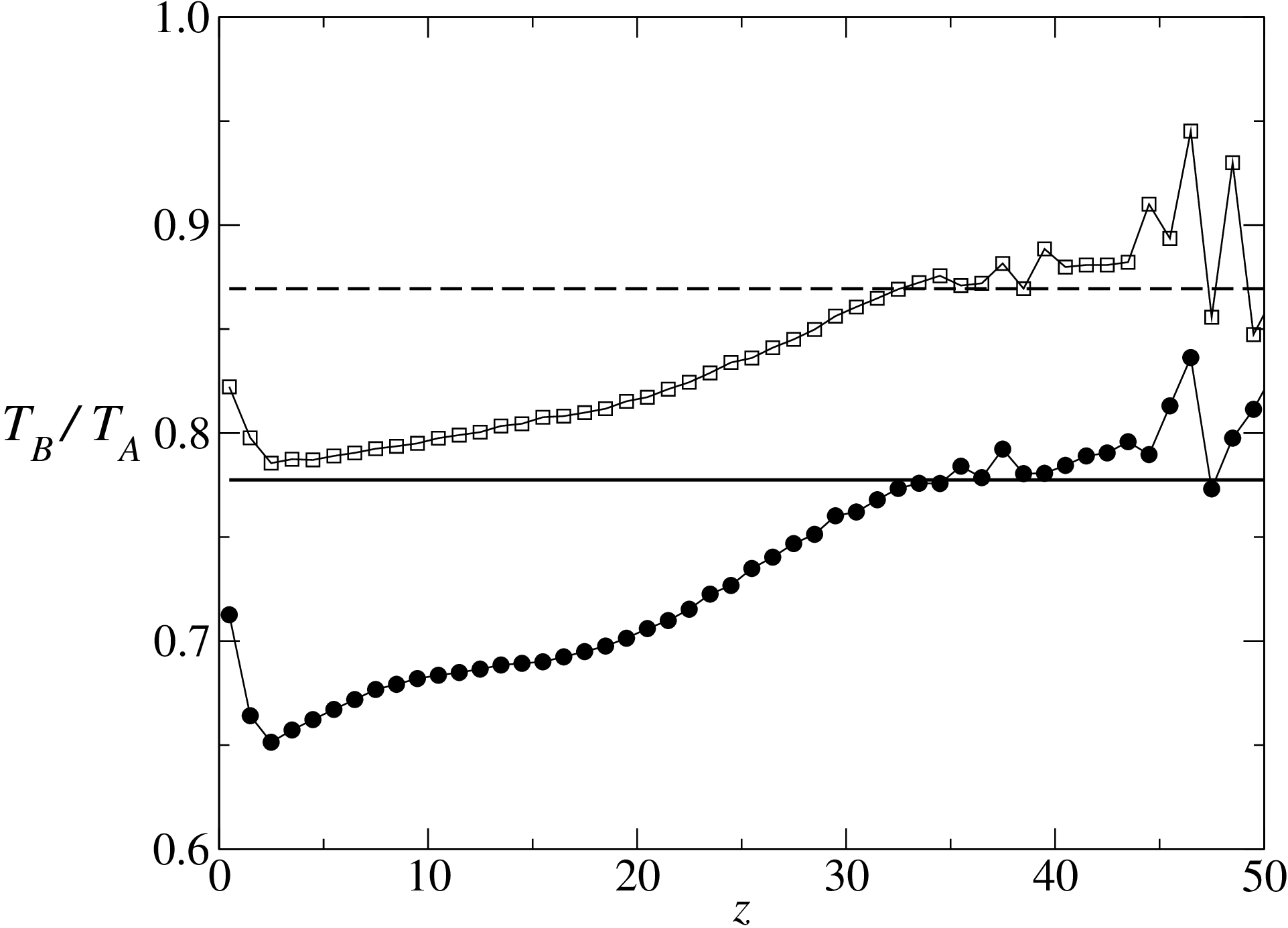}
\caption{Temperature ratios $T_B(z)/T_A(z)$ for the simulations
described in Fig. \ref{fig.tempprofiles}. Simulation results (solid circles) and
theoretical prediction in \cite{Brey} (solid line) for $N_A=670$, $N_B=10$,
$\alpha_A=0.98$, $\alpha_B=0.7$. Simulation results (open squares)
and
theoretical prediction in \cite{Brey} (dashed line) for A elastic,
$N_A=640$, $N_B=40$, and $\alpha_B=0.5$}
\label{fig.Tempratio}
\end{figure}

Figure \ref{fig.Tempratio} shows the temperature ratios $T_B(z)/T_A(z)$ for the simulations
described in Fig. \ref{fig.tempprofiles}. At low
densities it was found
experimentally \cite{Wildman,Menon} and by employing kinetic theory \cite{Brey}
that such temperature ratio must be constant. However, we find a non constant ratio
in the $z$ direction that only agrees with the result of \cite{Brey} at high
$z$.
Their prediction is  valid at low densities in the tracer limit,
conditions that are only reached in our case for large values of $z$.
In the second case, where A particles are elastic, equivalent predictions were given in
\cite{Martin,Trizac}.

In order to quantify the segregation, a series of simulations are performed with
$N_B$ ranging from 10 to 160, and $\alpha_B$ between 0.2 and 0.9. Larger values of $N_B$ or
smaller restitution coefficients lead to clustering as described in \cite{Meerson}.
For each simulation we compute the segregation parameter, defined as:
\beq 
\delta =1- 
 \int dz\, n_A(z)n_B(z)\bigg/ \sqrt{\textstyle\int dz\, n_A^2(z)\int dz\, n_B^2(z)}
\eeq
where the $n_A(z)$ and $n_B(z)$ are the local density, as plotted in Fig.
\ref{fig.densprofiles}.
The segregation parameter is bounded between 0 and 1.  The value $\delta=1$ 
corresponds to complete segregation, as $\delta$ equals 1 only if $n_A(z)$ and $n_B(z)$
do not overlap. On the contrary, $\delta=0$ means complete mixing, as this value
can only be obtained if $n_B(z)$ is
proportional to $n_A(z)$. 

\begin{figure}[htb]
\includegraphics[angle=0,clip=true,width=0.9\columnwidth,
keepaspectratio]{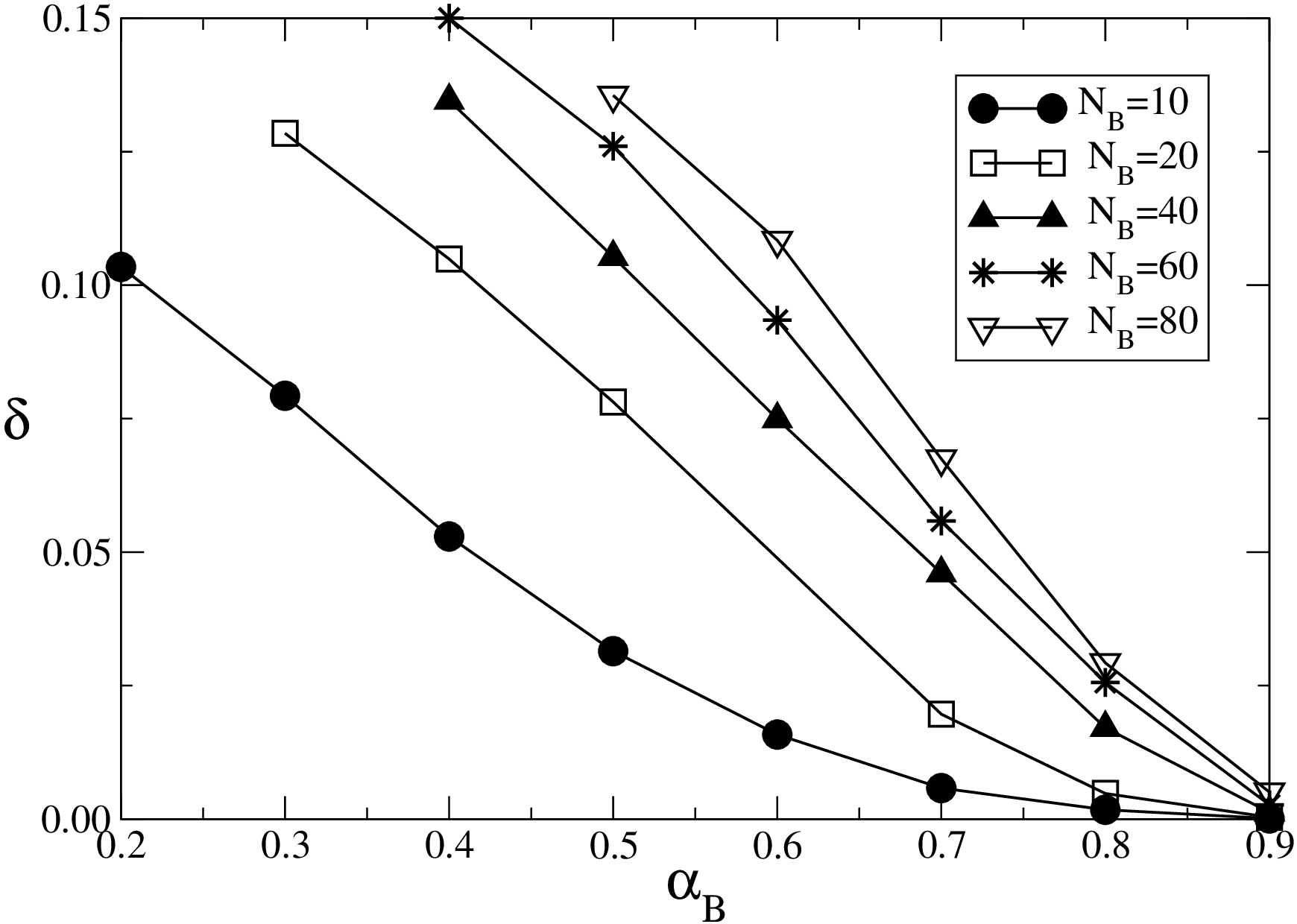}
\caption{Segregation parameter $\delta$ as a function of
$\alpha_B$. Different curves correspond to various
concentrations of $B$ particles. In the elastic limit $\alpha\to 1$ all curves 
coincide at $\delta=0$.}
\label{fig.delta}
\end{figure}

The results for $\delta$ are collected in Fig. \ref{fig.delta} where the quantity
$\delta$ is plotted versus the coefficient $\alpha_{B}$ for different values
of $N_B$. 
The fact that $\delta$ is always non vanishing
confirms that the segregation exists whenever the restitution coefficients are
different. Only in the case when $\alpha_B \to 1$ the quantity $\delta$
approaches 0, limit in which there is no segregation. Note that, for each
$\alpha_B$, $\delta$ increases with $N_B$. The results confirm that
segregation is not complete as $\delta$ never gets close to 1.
In addition, for each simulation, the center of mass of the A and B
species are computed, $Z_{A/B}$, finding that $Z_A>Z_B$ in all cases.

\begin{figure}[htb]
\includegraphics[angle=0,clip=true,width=0.9\columnwidth,
keepaspectratio]{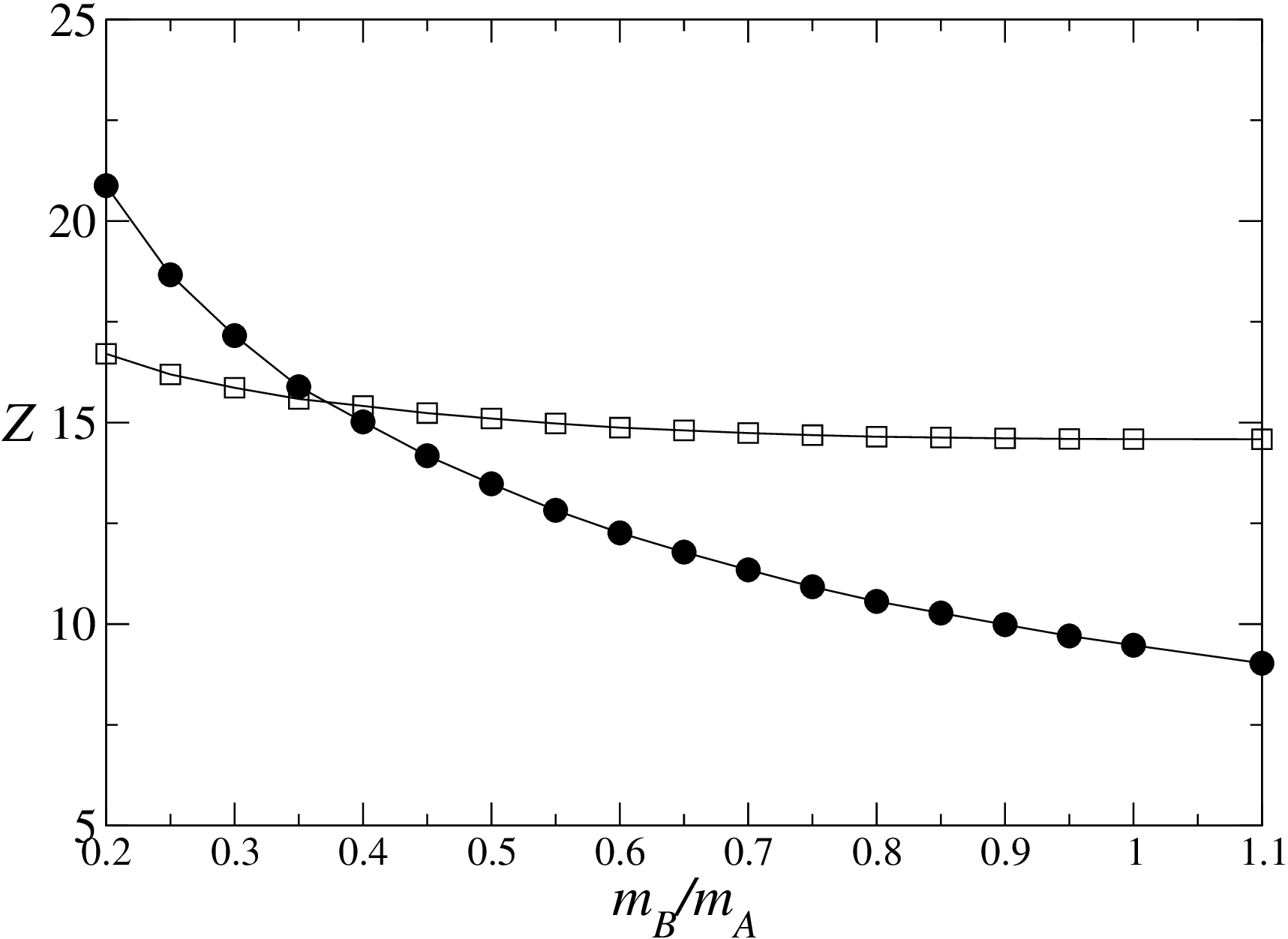}
\caption{Position of the center of mass of $A$ (open squares) and $B$
particles (solid circles) as a function of the relative mass of $B$ particles
$m_B/m_A$. 
Simulation parameters are fixed to $\alpha_A=1$, $\alpha_B=0.5$, $N_A=640$,
$N_B=40$, and $m_A(A\omega)^2=1$.}
\label{fig.varyingmB}
\end{figure}

It can be asked whether the observed segregation could compensate the
buoyancy force experienced by lighter $B$ particles. To verify this idea, 
a series of
simulations, keeping fixed $\alpha_A=1$, $\alpha_B=0.5$, $N_A=640$, $N_B=40$,
and $m_A(A\omega)^2=1$, but varying $m_B/m_A$ is performed. In each simulation,
the position of the center of masses of the $A$ and $B$ species are computed, $Z_A, Z_B$.
The results are presented in Fig. \ref{fig.varyingmB}. The inelastic particles
have lower center of mass if $m_B/m_A> 0.37$, and therefore sinking due to dissipation wins to the buoyancy force in this range. 
On the contrary, buoyancy force 
dominates if  $m_B/m_A < 0.37$ and inverse segregation is
obtained when $B$ particles are lighter than this threshold. The segregation parameter $\delta$ does not vanish for any value of $m_B/m_A$, indicating that there is no complete mixing even at the value of $m_B/m_A = 0.37$, where both center of masses coincide. The value of $\delta$, however, is minimum at this precise mass ratio. 

\section{Microstructure}

\begin{figure}[htb]
\includegraphics[angle=0,clip=true,width=0.9\columnwidth,
keepaspectratio]{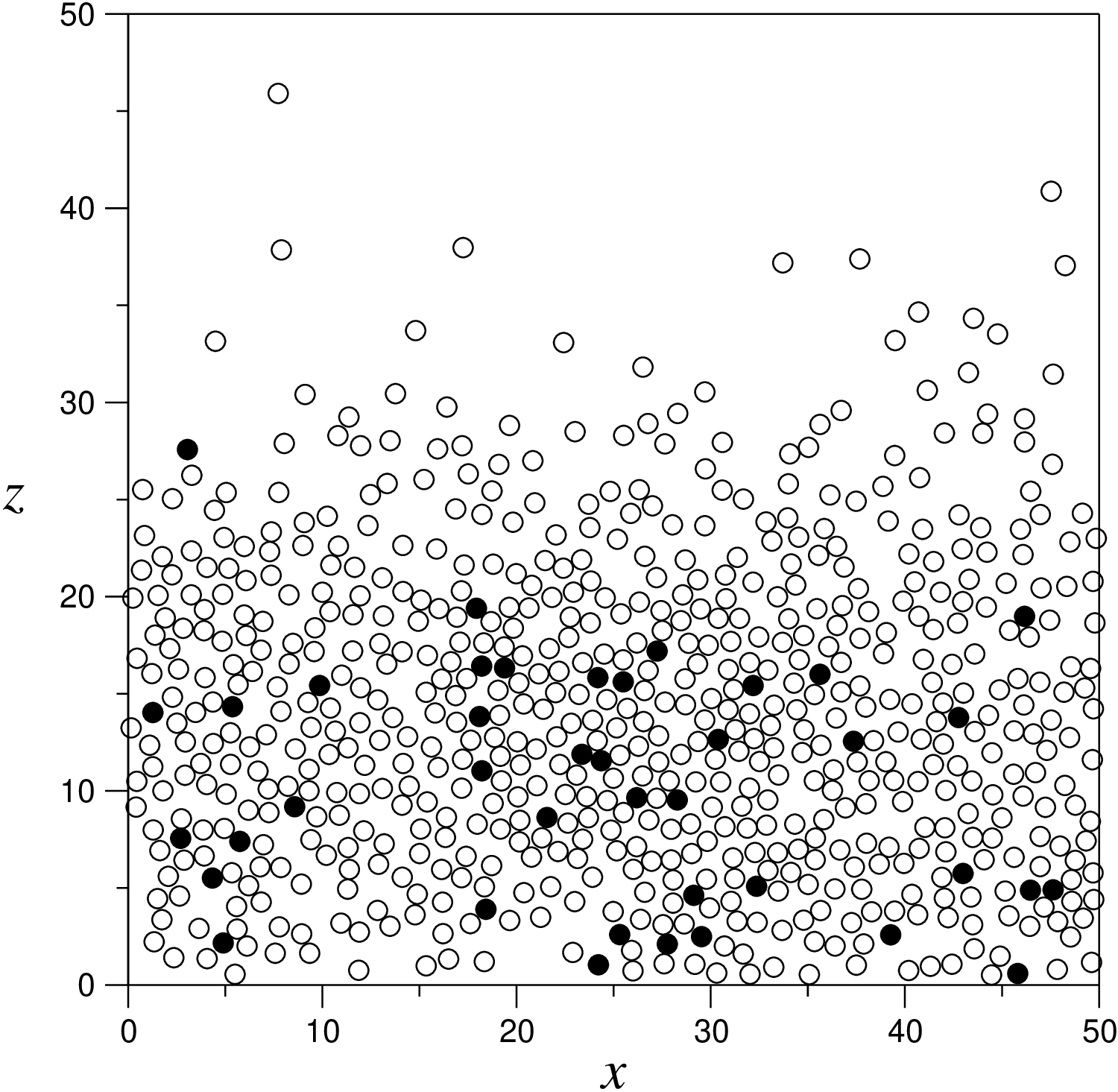}
\caption{Snapshot of the simulated system for $N_B=40$, $\alpha_B=0.5$.
A particles are white and B particles black. The border of some particles
exceed the lateral system size due to the periodic boundary conditions in this
direction.}
\label{fig.snapshot}
\end{figure}

A snapshot of the previously studied case, $\alpha_A=1, \alpha_B=0.5$ and $N_B=40$
is shown in Fig. \ref{fig.snapshot}.
The particles of type A are plotted as open circles,
and B are black symbols.
This snapshot suggest that, besides the macroscopic partial segregation
characterized by $\delta$, there is also a
micro-segregation, 
where B particles tend to be close to other B
particles, differently as if the particles where labeled A or B at
random. To decide if this observation is indeed true, we analyze
quantitatively the system computing the pair correlation functions.

As the system is inhomogeneous in the $z$ direction, the pair correlation
functions became
$z$-dependent. We chose to divide the
system in horizontal slabs of height $10\sigma$. For every pair of
particles, the center of mass is obtained and, according to it, their
relative distance is considered for the histogram of distances
associated to that particular slab. Furthermore, the pairs are classified according
to the type of particles involved. Finally, the histograms are normalized as
usual with the local density of each species in the slab, such that
at long distances the obtained pair correlation function approaches to unity.
This results in functions $g_{\mu\nu}^s(r)$, for the species $\mu,\nu
=\{A,B\}$, and the slab $s$.

In most of the studied cases, the region of density closest to homogeneous
corresponds to the second slab ($10\sigma\leq z<20\sigma$). In what
follows we will present results only for this slab and we will suppress
accordingly
the superscript in the pair correlation functions. The qualitative
properties for other slabs are similar to this particular slab, although the
effects are reduced because at higher slabs the densities are smaller.

\begin{figure}[htb]
\includegraphics[clip=true,width=\columnwidth,keepaspectratio]{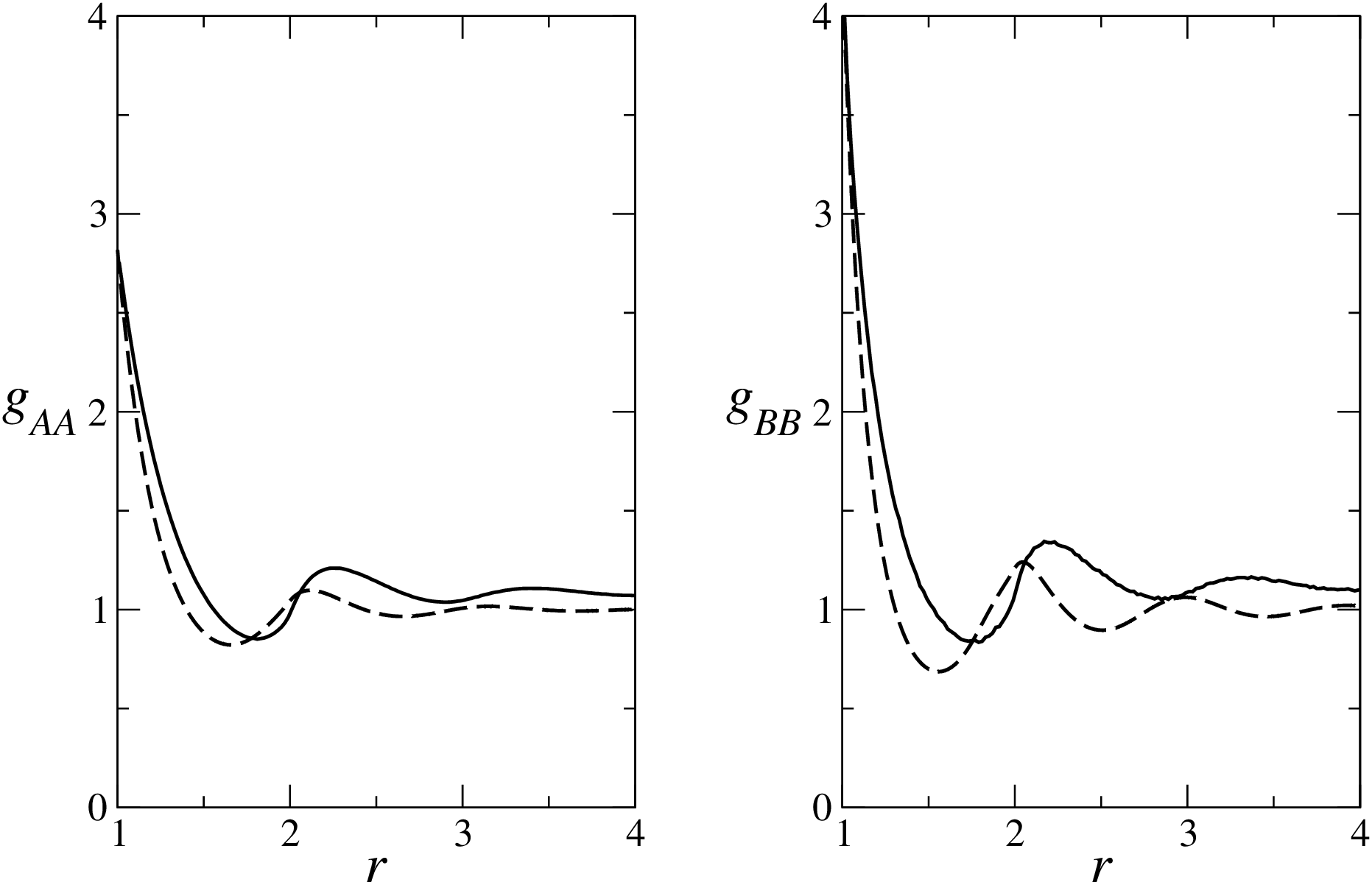}
\caption{Solid lines: radial distribution functions $g_{AA}(r)$, and
$g_{BB}(r)$. $N_B=100$, $\alpha_B=0.8$. Dashed lines: equilibrium radial
distribution function given in Ref.~\cite{Kolafa}, with the density fitted to
give the same value of $g_{AA}$ or $g_{BB}$ at contact (left $n_{AA}=0.636$,
right $n_{BB}=0.823$) }
\label{fig.gr}
\end{figure}

Figure \ref{fig.gr} shows the density correlation function for the pairs AA  and
BB for $N_A=580$, $N_B=100$, $\alpha_B=0.8$,  conditions in which the system
develops an average density of $n\simeq 0.56$. Correlation of distinct particles AB have
intermediate properties between AA and BB. The main noticeable feature of these correlation
functions is that the first and second peak of the BB function is much larger than the AA one. In other
words, the large values of $g_{BB}$ at contact means that the are more B-B pairs in the system
that in a configuration where A and B particles are labeled at random.
This excess number of B-B pairs could be guessed from the figure \ref{fig.snapshot}, where one can
easily locate 4 B-B pairs.

The structure of the correlation functions in our inelastic system
could resemble  the radial distribution of an elastic fluid at a {\em selected}
higher density. We tried to exploit this idea by comparing the
granular distribution function with an elastic one by choosing an appropriate density.
The way to select the density is by adjusting the value of the pair distribution
function {\em at contact}.  We take this criterion inspired by the kinetic theory of dense
fluids, where the pair distribution function at contact is used to improve the Boltzmann
equation  including certain correlations. With such procedure we find different fitting
densities for AA and BB pairs, being their values: $n_{BB}=0.823$ and
$n_{AA}=0.636$.
Surprisingly, none of them (not even the elastic) agree with the average density in the
system, $n\simeq 0.56$. This means that the dissipative particles, type B, are
able to modify the structural properties of the A particles, which we have chosen
to be elastic. The fitted elastic pair correlation functions are plotted as dashed lines in
Fig. \ref{fig.gr}. Despite of the fitting procedure, large discrepancies are observed between our correlation
functions and the elastic ones. Other fitting procedures could have been selected, but none of
them produce an agreement of the correlation functions in all its range.

The inelastic
correlation function shows a strong enhancement of the first peak, followed by a second,
and at most a third peak, reflecting a short range structure of few particle diameters.
However, the location of the peaks differs form the elastic
correlation function, indicating changes in the microstructure.
The fast decay of the correlation function in inelastic systems has been observed in
homogeneously heated inelastic gases as well \cite{Ignacio}. This analysis
confirms the existence of a microsegregation of particles of type B.

In addition, the pair correlation function at contact $\chi=g(\sigma^+)$ is computed
for the pairs AA, AB, and BB.
The values of $\chi$ grow when increasing $N_B$ and decreasing
$\alpha_B$, obeying that $\chi_{BB}>\chi_{AB}> \chi_{AA}$.
Finally let us remark that $\chi_{BB}$ can reach very high values. For instance, at $N_B=160$ and $\alpha_B=0.5$,
$\chi_B\simeq 60$, despite that the average total density is only  $n\simeq 0.5$.

It has been proposed \cite{Lutsko}, based on an
Enskog-like kinetic model that, due to the smaller outgoing velocity
after a collision, particles stay longer in the close vicinity, when
the restitution coefficient is less than one. These arguments leads to an enhancement
of the pair correlation function in terms of the restitution coefficient $\alpha$, as
\beq
\chi(\alpha) = \frac{1+\alpha}{2\alpha}\chi_0.  \label{chi}
\eeq
Here $\chi_0$ is the pair correlation function of elastic particles at
the same density, that are given by the Verlet-Levesque \cite{LV} and Carnahan-Starling
\cite{CS} factors for 2 and 3 dimensions.

\begin{figure}[htb]
\includegraphics[angle=0,clip=true,width=0.9\columnwidth,
keepaspectratio]{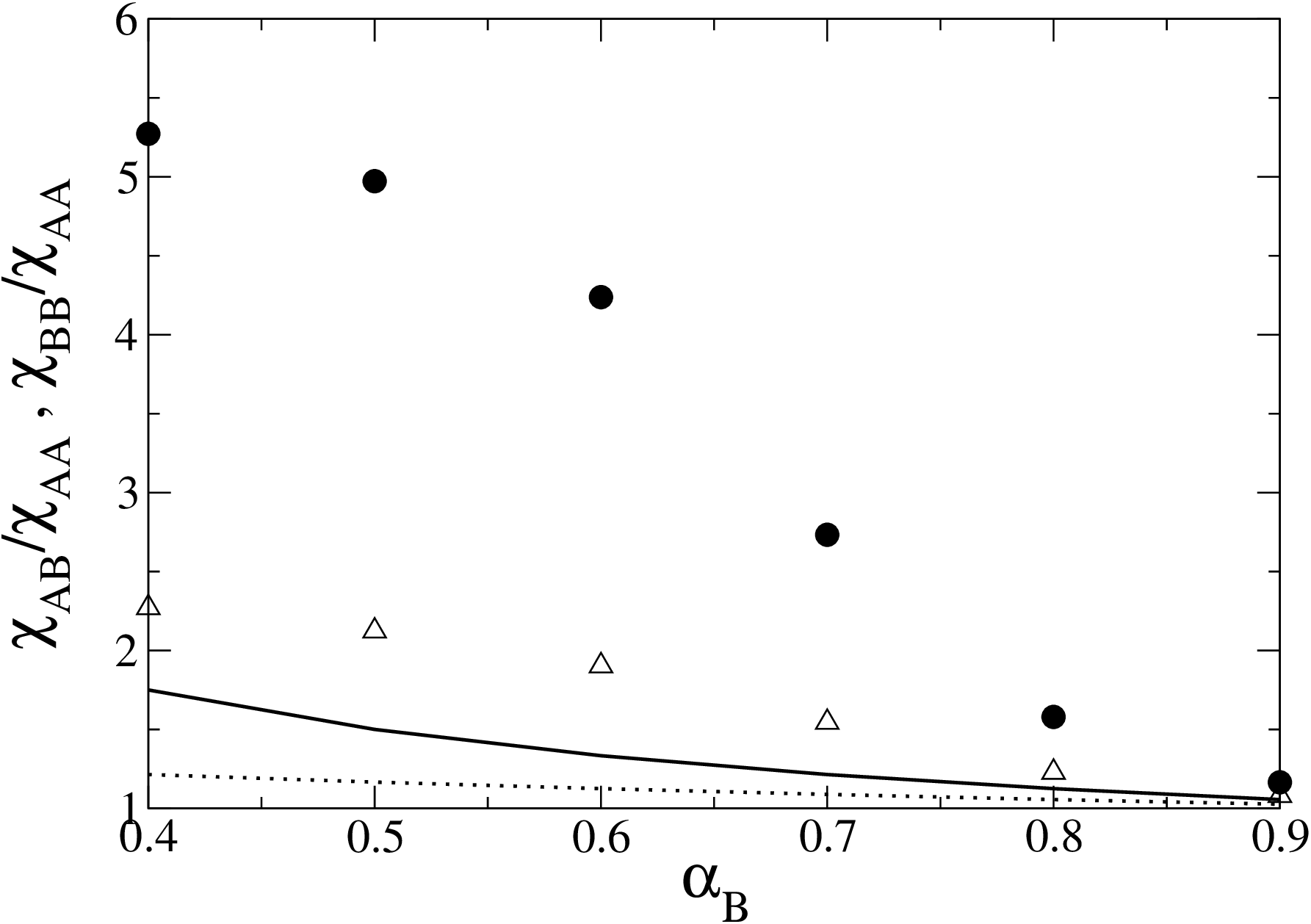}
\caption{Contact pair correlation function quotients as a
function of the restitution coefficient. $\chi_{BB}/\chi_{AA}$ results
from simulations (filled circles) and the Enskog theoretical value
(solid line); and $\chi_{AB}/\chi_{AA}$ results
from simulations (open triangles) and the Enskog theoretical value
(dotted line). Here $N_B=120$.}
\label{fig.cuocientechi}
\end{figure}

To test if this prediction is valid we have plotted in Fig. \ref{fig.cuocientechi} the ratios
$\chi_{BB}/\chi_{AA}$ against $\alpha_B$ and $\chi_{AB}/\chi_{AA}$ against
$\alpha_{AB}$ and compared them with the factors $(1+\alpha_B)/(2\alpha_B)$
and $(1+\alpha_{AB})/(2\alpha_{AB})$ respectively. Dividing by $\chi_{AA}$ we get rid of
the $\chi_0$ factor in (\ref{chi}).
The figure
indicates that the Enskog model is not enough to describe the high values of
the pair correlation when one inelastic particle is present, as was
already quoted by its author \cite{Lutsko}. A possible
origin of this effect are recollisions, that are not taken into account in
Enskog's theory, a phenomenon that is pronounced in granular systems.
Inelastic particles, after a collision, separate at a slower speed,
increasing the possibility of having a collision with a third particle,
approaching it again the original pair. In some way, inelasticity
increases the so-called {\em cage effect} in liquids.

\begin{figure}[htb]
\includegraphics[angle=0,clip=true,width=0.9\columnwidth,
keepaspectratio]{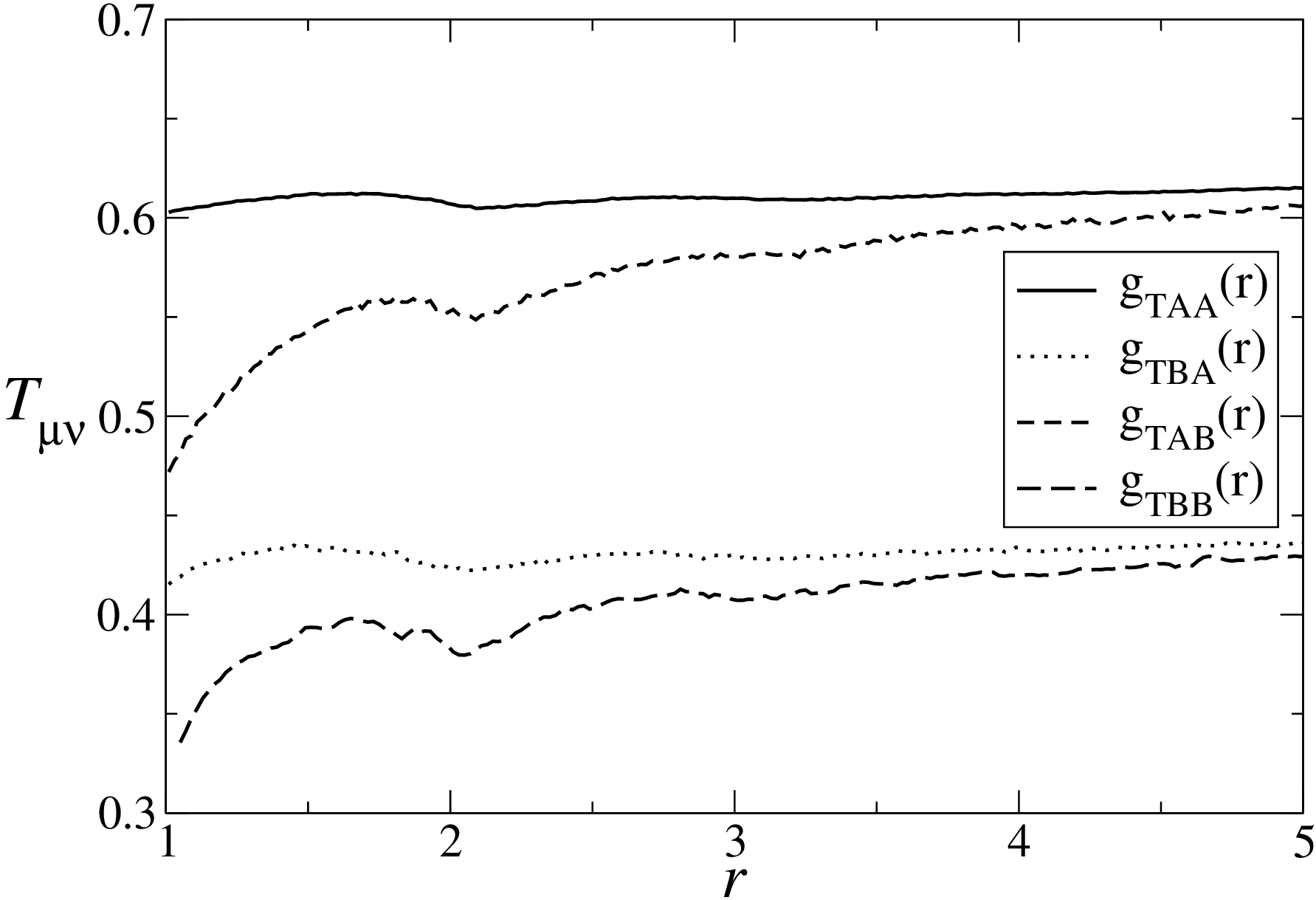}
\caption{Radial temperatures $T_{\mu\nu}$ of a particle of type
$\mu$ around a particle of type $\nu$. $N_B=40$, $\alpha_B=0.5$.}
\label{fig.gTr}
\end{figure}

Besides density correlations induced by B particles, they also create a local decrease
of temperature because their collisions are inelastic.  We have computed the radial temperature
of particles of type $\mu$ located at a distance $r$ of a $\nu$ particle:
$T_{\mu\nu}(r)$. Figure \ref{fig.gTr} shows these radial temperature functions.
Their asymptotic values are the average temperatures of A and B particles in the slab $10\sigma < r < 20\sigma$.
Besides that B particles are colder than A particles, they produce a local decrease
of temperature both for A and B particles. This effect is more pronounced when
increasing the inelasticity but almost independent of the concentration of B particles,
whose main effect is to reduce globally the
temperature. As before this is a local effect that extends for
about $2\sigma$.

\section{Origin of the global segregation}

So far we have argued along the paper that B particles locally modify the structure
of B particles (inelastic) but also of A particles (elastic).
For instance, Fig. \ref{fig.gr} shows that
the density around B particles is higher than around A
particles. The excess mass around B particles with respect A particles is defined as,
\beq\label{excessmass}
\delta m = \int_\sigma^{\infty} dr\, 2\pi r(\rho_B(r)-\rho_A(r)),
\eeq
where $\rho_\mu(r)$ is the total density around a particle of type $\mu$. The excess mass is positive
for all values of $N_A, N_B$ and restitution coefficient $\alpha_B$.
Similarly, the local temperature around a B particle is lower than the local
temperature around a A particle (see Fig. \ref{fig.gTr}).

These two results imply that around a B particle it is developed a dense
and cold region, that we call a {\em cold droplet}. The characteristic size of the droplet
is microscopic; in agreement with the plots of Sect IV, must be about 2-3 $\sigma$.
The development of a cold
droplet around each B particle increases their effective mass and also, due to the lower
temperature, decreases the pressure. This mechanism could be similar
(at the local level) to the clustering instability described in
\cite{Goldhirsch}. In our case the unstable process does not continue as the
energy injection due to the vibration breaks the droplet, that must be seen as dynamical object,
that forms and evanesces continuously.
Once the cold droplet is formed around B the buoyancy force in weaker than its effective weight
and therefore there is a net force that
tends to sink the cluster and consequently the B particle.

This continuously sink of the B particle is the responsible of the macroscopic
segregation described in Sect. III.

\section{Summary and Conclusions}

The main conclusion of the present paper is that different restitution coefficients alone
create  segregation in a binary mixture vertically vibrated. The restitution coefficients
must be considered, in addition to the usual material properties (mass ratio and diameter ratio)
in order to describe accurately the segregation.
The effect of the inelasticity is such that the most inelastic particles sink to
the bottom of the 
container while the less inelastic ones raise to the top. Segregation is not
complete, however, but 
only partial. The density profiles of each species shows a maximum which is
located in a different 
position depending on the inelasticities. Concerning the temperatures, most
dissipative particles have 
a lower temperature than the most elastic ones, both having the characteristic
shape of vibrofluidized 
system (fast cooling away from the moving boundary followed by a heating that
grows linearly with the distance).

The segregation effects presented here also appear by vibrating a
mixture of elastic and inelastic particles. Again inelastic particles migrate to
the bottom of the container and elastic one prefer the upper part. The
temperature distribution also
looks similar to the full inelastic mixture.

Besides the macroscopic segregation there is also segregation at the microscopic scale.
With the word microscopic we refer to properties at distances of few particles diameters.
In our case we find a
notorious increase of the probability of finding two inelastic particles together as compared with the less
inelastic or elastic ones. This enhancement cannot be described by only considering kinematic properties.
Our guess is that dynamic correlations are required to properly describe these correlations.

Finally, we propose a mesoscopic explanation to the segregation: the most inelastic particles induce
locally a region of high density and low temperature, resembling a cold droplet that falls in a
gravitational field. The droplet is created by the dissipation in a way that resembles the clustering
instability of the granular gases.

The effect of the different restitution coefficients may act in an opposite direction as the usual Brazil
nut effect. For instance, consider a vibrated granular fluid and insert a large (or light) intruder than tends to
move upward. If the intruder is very dissipative, it will move downwards, as we have described in this paper.
Which is the final position? Which force does finally win? Partial answer is presented in Fig.~\ref{fig.varyingmB}, where we show the competition between the buoyancy force and the sinking effect due to dissipation. 
Moreover, could two large an inelastic intruders come
together by the effect of inelasticity alone, as shown in the Sect. IV of the present paper? Further research is needed in order to answer these questions.

\begin{acknowledgments}
We want to thank J.M.R. Parrondo for very useful comments.
R.B. is supported by the Spanish Projects MOSAICO,
FIS2004-271 and UCM/PR34/07-15859. The research is supported by {\em Fondecyt} grants
1061112, 1070958, and 7070301 and {\em Fondap} grant 11980002.
\end{acknowledgments}


\begin{thebibliography}{99}
\bibitem{Rosato} A. Rosato, K.J. Strandburg, F. Prinz, and R.H. Swendsen,
Phys. Rev. Lett. {\bf 58}, 1038 (1987).

\bibitem{Breu}A. P. J. Breu, H.-M. Ensner, C. A. Kruelle, and I. Rehberg
Phys. Rev. Lett. {\bf 90}, 014302 (2003).

\bibitem{Kudrolli} A. Kudrolli, Rep. Prog. Phys. {\bf 67}, 209 (2004).

\bibitem{RuizSuarez} D. A. Huerta and J. C. Ruiz-Su\'arez,
Phys. Rev. Lett. {\bf 92}, 114301 (2004)
Erratum: {\em ibid} {\bf 93}, 069901(E) (2004)

\bibitem{Schroter} 
M. Schr\"oter, S. Ulrich, J. Kreft, J.B. Swift, and H.L.
Swinney, Phys. Rev. E {\bf 74}, 011307 (2006).

\bibitem{Khakhar} D.V. Kharkar, J.J. McCarthy and J.M. Ottino, CHAOS {\bf 9}, 594 (1999).

\bibitem{Mullin}P. M. Reis and T.  Mullin,
Phys. Rev. Lett. {\bf 89}, 244301 (2002).

\bibitem{Ciamarra}M. Pica Ciamarra, A. Coniglio, and M. Nicodemi, Phys. Rev.
Lett. {\bf 94}, 188001 (2005)


\bibitem{Schnautz} T.~Schnautz, R. Brito, C.A. Kruelle, and I. Rehberg, Phys.
Rev. Lett. {\bf 95}, 028001 (2005).

\bibitem{Makse} H. A. Makse, S. Havlin, P. R. King, and H. E. Stanley, Nature
{\bf 386}, 379 (1997).

\bibitem{Hill} K. M. Hill and J. Kakalios, Phys. Rev. E {\bf 49}, R3610, (1994).

\bibitem{King}
D. A. Sanders, M. R. Swift, R. M. Bowley, and P. J. King, Phys. Rev.
Lett. {\bf  93}, 208002 (2004).

\bibitem{King2} L. T. Lui, Michael R. Swift,
R. M. Bowley, and P. J. King, Phys. Rev. E {\bf 75}, 051303 (2007)

\bibitem{Kondic} L. Kondic, R.R. Hartley, S.G.K. Tennakoon, B. Painter, 
and R.P. Behringer, Europhys. Lett, {\bf 61}, 742 (2003).

\bibitem{Serero} D. Serero, I Goldhirsch, S.H. Noskowick and M.-L. Tan, J. Fluid Mech.
{\bf 554}, 237 (2006). The case of different restitution coefficients is
treated in Section 5.1.

\bibitem{Brey} J.J. Brey, M.J. Ruiz-Montero and F. Moreno, Phys. Rev. E {\bf
73}, 031301 (2006).

\bibitem{Wildman} R.D. Wildman and D.J. Parker,
Phys. Rev. Lett. {\bf 88}, 064301 (2002).

\bibitem{Menon} K. Feitosa and N. Menon, Phys. Rev. Lett. {\bf 88}, 198301 (2002).

\bibitem{Paolotti} D. Paolotti, C. Cattuto, U. Marini Bettolo Marconi, A.
Puglisi, Granular Matter {\bf 5}, 75(2003).

\bibitem{Garzo} V. Garz\'o, Europhys. Lett, {\bf 75}, 521 (2006). V. Garz\'o, cond-mat/0803.2588. 

\bibitem{Soto} R. Soto, Phys. Rev. E {\bf 69} 061305 (2004).

\bibitem{Grossman} E. L. Grossman, T. Zhou, and E. Ben-Naim, Phys. Rev. E
{\bf 55}, 4200 (1997).

\bibitem{RamirezSoto} R. Ram\'{\i}rez and R. Soto, Physica A {\bf 322} 73
(2003)

\bibitem{Martin} Ph. A. Martin and J. Piasecki, Europhys. Lett., {\bf 46}, 613 (1999).

\bibitem{Trizac} A. Barrat, E. Trizac, Granular Matter {\bf 4}, 57 (2002).

\bibitem{Meerson} B.  Meerson, T. P\"oschel, and Y. Bromberg, Phys. Rev. Lett.
{\bf 91}, 024301 (2003). 

\bibitem{Ignacio}
I. Pagonabarraga, E. Trizac, T. P. van Noije, and M. H. Ernst,
Phys. Rev. E {\bf 65}, 011303 (2002).

\bibitem{Kolafa} J. Kolafa, S. Lab\'{\i}k, and A. Malijevsk\'y, Phys.
Chem. Chem. Phys. \textbf{6}, 2335 (2004). See also
http://www.vscht.cz/fcs/software/hsmd/ for molecular dynamics results of
$g(r)$.

\bibitem{Lutsko} J. F. Lutsko. Phys. Rev. E {\bf 63}, 061211 (2001).

\bibitem{LV} L. Verlet and D. Levesque, {\em Mol. Phys.} {\bf 46}, 969 (1982).

\bibitem{CS} N. F. Carnahan and K. E. Starling, J. Chem. Phys. {\bf 51}, 635
(1969).

\bibitem{Goldhirsch} I. Goldhirsch and G. Zanetti, Phys. Rev. Lett. {\bf
70}, 1619 (1993).
\end{thebibliography}
\end{document}